\documentclass{PoS}

\title{Preparation to the CMB PLANCK data analysis, estimation of the contamination due to the galactic polarized emissions}

\ShortTitle{Galactic emission models}

\author{\speaker{ L. Fauvet}\\ LPSC, Universit\'e Joseph Fourier Grenoble 1, CNRS/IN2P3, Institut National Polytechnique de Grenoble, 53 avenue des Martyrs,\\
38026 Grenoble cedex, France\\
        E-mail: \email{fauvet@lpsc.in2p3.fr}}

\author{J.-F. Mac\'ias P\'erez \\ LPSC, Universit\'e Joseph Fourier Grenoble 1, CNRS/IN2P3, Institut National Polytechnique de Grenoble, 53 avenue des Martyrs,\\
38026 Grenoble cedex, France\\
        E-mail: \email{macias@lpsc.in2p3.fr}}

\abstract{This work is point of the preparation to the analysis of
  the PLANCK satellite data. The PLANCK satellite is an ESA mission which has been launched the 14th of may 2009 and
is dedicaced to the measurement of the Cosmic Microwave Background
(CMB) in temperature and polarization. The presence of diffuse
Galactic polarized emissions disturb the measurement of the CMB
anisotropies, in particular in polarization. Therefore a precise
knowledge of these emissions is needed to obtain the level of accuracy
required for PLANCK. In this context, we have developed and
implemented a coherent 3D model of the two mains polarized Galactic
emissions : synchrotron and thermal dust. We have compared these models to
preexisting data: the 23 GHz band of the WMAP data, the 353 GHz Archeops data and the 408 MHz all-sky continuum
survey. We extrapolate these models to the frequencies where the CMB
dominates and we are
able to estimate the contribution of polarized foreground emissions to
the polarized CMB emission measured with PLANCK.}

\FullConference{International Workshop on Cosmic Structure and Evolution - Cosmology2009,\\
                September 23-25, 2009\\
                Bielefeld, Germany}

\begin{document}

\section{Introduction}


The PLANCK satellite, currently in flight, should give the more
accurate measurement of the anisotropies of the CMB in temperature and
polarization with a sensitivity of $2 \mu K$ and an angular resolution
of 5 arcmin~\cite{bluebook}. In particular its estimation of the
BB-modes should set an upper limit on the tensor-scalar ratio
(expected at 0.1~\cite{efstathiou}). The knowledge of this ratio
should confirm the existence of primordial gravitational waves
generated during the inflation and would set the energy scale of the
inflation~\cite{lyth} and provide constraint on inflationnary models~\cite{baumann}. In order to obtain its optimal sensitivity it is required to estimate the foreground emissions and
the residual contamination due to these foreground emissions on the
CMB signal. Indeed for the full sky these emissions have the same order
of magnitude than the CMB in temperature and dominate by a factor of 10 in
polarization~\cite{bluebook}. The principal polarized Galactic microwave emissions come
from 2 effects : thermal dust emission and synchrotron emission. The
synchrotron has already been measured by the 408 MHz all-sky continuum survey~\cite{haslam}, by Leider between 408 MHz and 1.4 GHz~\cite{wolleben}, by Parkes at 2.4 GHz~\cite{duncan1999}, by the MGLS {\it Medium Galactic Latitude Survey} at
1.4 GHz~\cite{uyaniker} and by the satellite WMAP {\it Wilkinson Microwave Anisotropies Probe} (see e.g.~\cite{hinshaw}). The synchrotron emission is due to ultrarelativist
electrons spiraling in a large-scale magnetic field and is dominant at
low frequencies. The dust thermal
emission which have already been well constrained by IRAS~\cite{schlegel}, COBE-FIRAS~\cite{boulanger} and
Archeops~\cite{macias,benoit} is due to dust grains
which interact with the Galactic magnetic field and emit a polarized
submillimetric radiation~\cite{boulanger} and dominates at high frequencies. The polarization of these
two radiation is orthogonal to the field lines. To obtain a realistic model of
these emissions we propose models based on a 3D modelling of the
Galactic magnetic field and of the matter density in the
Galaxy. The models are optimized using preexisting data and then are used to estimate the bias due to
these emissions on the CMB measurement.

\section{3d modelling of the Galaxy}
\label{sec:model}

\indent A polarized emission is described by the Stokes
parameters I, Q and U~\cite{kosowsky}. For the polarized foreground
emissions integrating along the line of sight we obtain, for synchrotron~\cite{ribicki}:
\begin{eqnarray}
\label{eq:map_sync}
\centering
 I_s &=& I_{\mathrm{Has}} \left(\frac{\nu_s}{0,408}\right)^{\beta_s},\\
Q_s &=& I_{\mathrm{Has}}
\left(\frac{\nu_s}{0,408}\right)^{\beta_s}\frac{\int \cos(2\gamma)p_s n_e\left(B_l^2 + B_t^2 \right)}{\int n_e\left(B_l^2 + B_t^2 \right)} ,\\
U_s &=& I_{\mathrm{Has}}
\left(\frac{\nu_s}{353}\right)^{\beta_s}\frac{\int \sin(2\gamma)p_s n_e\left(B_l^2 + B_t^2 \right)}{\int n_e\left(B_l^2 + B_t^2 \right)},
\end{eqnarray}

\noindent where $B_n$, $B_l$ and $B_t$ are the magnetic field components
along, longitudinal and transverse to the ligne of sight. $p_s$ is the
polarization fraction set to 75\%~\cite{ribicki}. $I_{Has}$ is a
template temperature map obtained from the 408 MHz all-sky continuum survey~\cite{haslam}. The maps are extrapolated to the Planck
frequencies using the spectral index $\beta_s$ which is a free parameter of the model.

For the thermal dust emission : \\
\begin{eqnarray}
\centering
 I_d &=& I_{sfd} \left( \frac{\nu_d}{353} \right)^{\beta_d},\\
Q_d &=& I_{sfd} \left( \frac{\nu_d}{353}\right)^{\beta_d} \int n_d\frac{\cos(2 \gamma) \sin^2(\alpha) f_{\mathrm{norm}}p_d}{n_d} ,\\
U_d &=& I_{sfd} \left(\frac{\nu_d}{353}\right)^{\beta_d}  \int n_d \frac{\sin(2 \gamma) \sin^2(\alpha) f_{\mathrm{norm}}p_d}{ \int n_d} ,
\end{eqnarray}

\noindent where the polarization fraction $p_d$ is set to 10 \%~\cite{ponthieu2005},
$\beta_d$ is the spectral index (set at 2.0) and $f_{norm}$ is an
empiric factor, fit to the Archeops data. The
$I_{sfd}$ map is the model 8 of~\cite{finkbeiner}.\\ 

\indent The models are based on a exponential distribution of
relativistic electrons on the Galactic disk
following~\cite{drimmel} where the radial scale $h_r$ is a free
parameter. The distribution of dust grains $n_d$ is
also choose exponential~\cite{benoit}. The Galactic magnetic field is composed of two
parts: a regular component and a turbulent component. The regular
component is based on the WMAP team model~\cite{page} which is close to a
logarithmic spiral to reproduce the shape of the spiral
arms~\cite{han2006,sofue}. The pitch angle $p$ between two arms is a free
parameter of the model. The turbulent component is described by
a law of Kolmogorov~\cite{han2006,han2004} spectrum of relative amplitude $A_{turb}$.

\section{Comparison to data}
\label{sec:test}

\begin{figure}
\centering
\includegraphics[height=8cm,width=6cm]{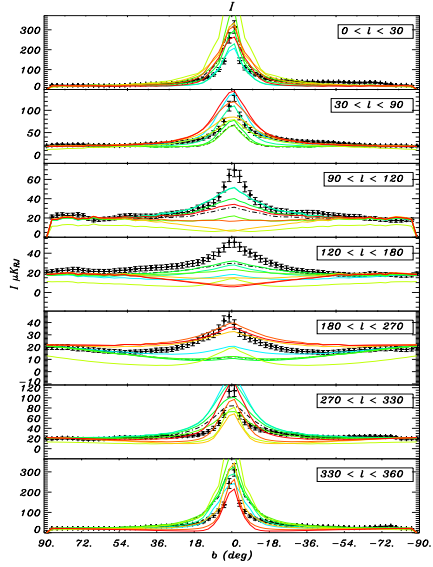}\caption{Galactic profiles in temperature at 408 MHz built using Haslam data in black and our synchrotron emission model for various values of the pitch angle $p$ {\it (form green to red)}.\label{fig:gal_has}}
\end{figure}

\indent We computed Galactic profiles in temperature and polarization
for various bands of longitude and latitude and various values of the free
parameters. In order to optimize these 3D models we compare them to
Galactic profiles computed from preexisting data using a $\chi^2$
test. For the synchrotron emission in temperature, we use the 408 MHz all-sky
continuum survey~\cite{haslam} as shown on
Figure~\ref{fig:gal_has}. In polarization we compared to the K-band
WMAP 5 years data. Thermal dust emission model is optimized using the
polarized Archeops data~\cite{benoit} at 353 GHz.

\indent The best fit parameters for the 3D model in polarization
are given in the Table~\ref{tab:param}. The results are consistent for
the 3 sets of data, and in particular we obtain compatible results
for the synchrotron and thermal dust emission models. $A_{turb}$ is
not strongly constrained but its range of best fit value is compatible
with previous results~\cite{sun,dusta,han2004}.
$h_r$ is badly constrained as was already the case in Sun {\it et al}~\cite{sun}. The
best fit value of the pitch angle $p$ is compatible with results
obtained by other study~\cite{sun,page}. 
The best fit value for the spectral index of the synchrotron
emission is lower than value found by~\cite{sun,page} but it is
probably due to the choice of normalisation using the 408 MHz template. \\

\begin{table}[h]
\begin{center}
\caption{Best fit parameters for synchrotron and thermal dust emission models and $3\sigma$ confidence levels for the best fitting model.\label{tab:param}}
\vspace{0.3cm}

\begin{tabular}{|c|c|c|c|c|c|} \hline
$$  & $ p (deg)$& $A_{turb} $  & $h_r$  &    $\beta_s$  & $\chi^2_{min}$  \\\hline
$WMAP$ & $ -30.0^{+40.0}_{-30.0}$ & $< 1.25$ (95.4 \% CL) & $ <20$ (95.4 \%
CL) &  $-3.4^{+0.1}_{-0.8}$ & $5.72$     \\\hline
$HASLAM$ & $ -20.0^{+60.0}_{-50.0}$   & $< 1.0$ (95.4 \% CL) &  $ 4.0^{+16.0}_{-3.0} $ & $\emptyset$ & $5.81$ \\\hline
$ARCHEOPS$ & $ -20^{+80}_{-50}$   & $ < 2.25 (95.4 \% CL)$ & $\emptyset$ & $\emptyset$ &  $ 1.98$          \\\hline

\end{tabular}
\end{center}
\end{table}

\begin{figure}
\centering
\includegraphics[height=9cm,width=13cm]{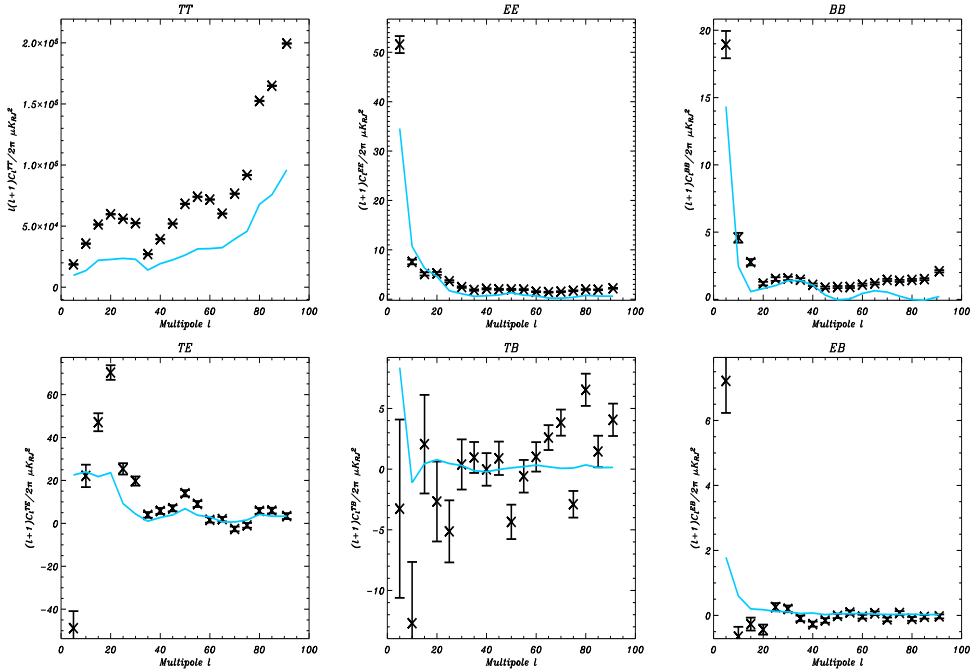}\caption{Clockwise from top left : power spectra $C^{TT}_l$, $C^{EE}_l$, $C^{BB}_l$, $C^{TE}_l$, $C^{TB}_l$, $C^{EB}_l$ at 23 GHz built with the WMAP 5 years data \emph{(black)} and the model of synchrotron emission with BSS magnetic field for the best fit model parameters \emph{(green to red for a spectral index between -3.2 and -3.4)}, applying a Galactic cut $|b|<5^{\circ}$.\label{spec_wmap_gp}}
\end{figure}

\begin{figure}
\centering
\includegraphics[angle=90,height=9cm,width=13cm]{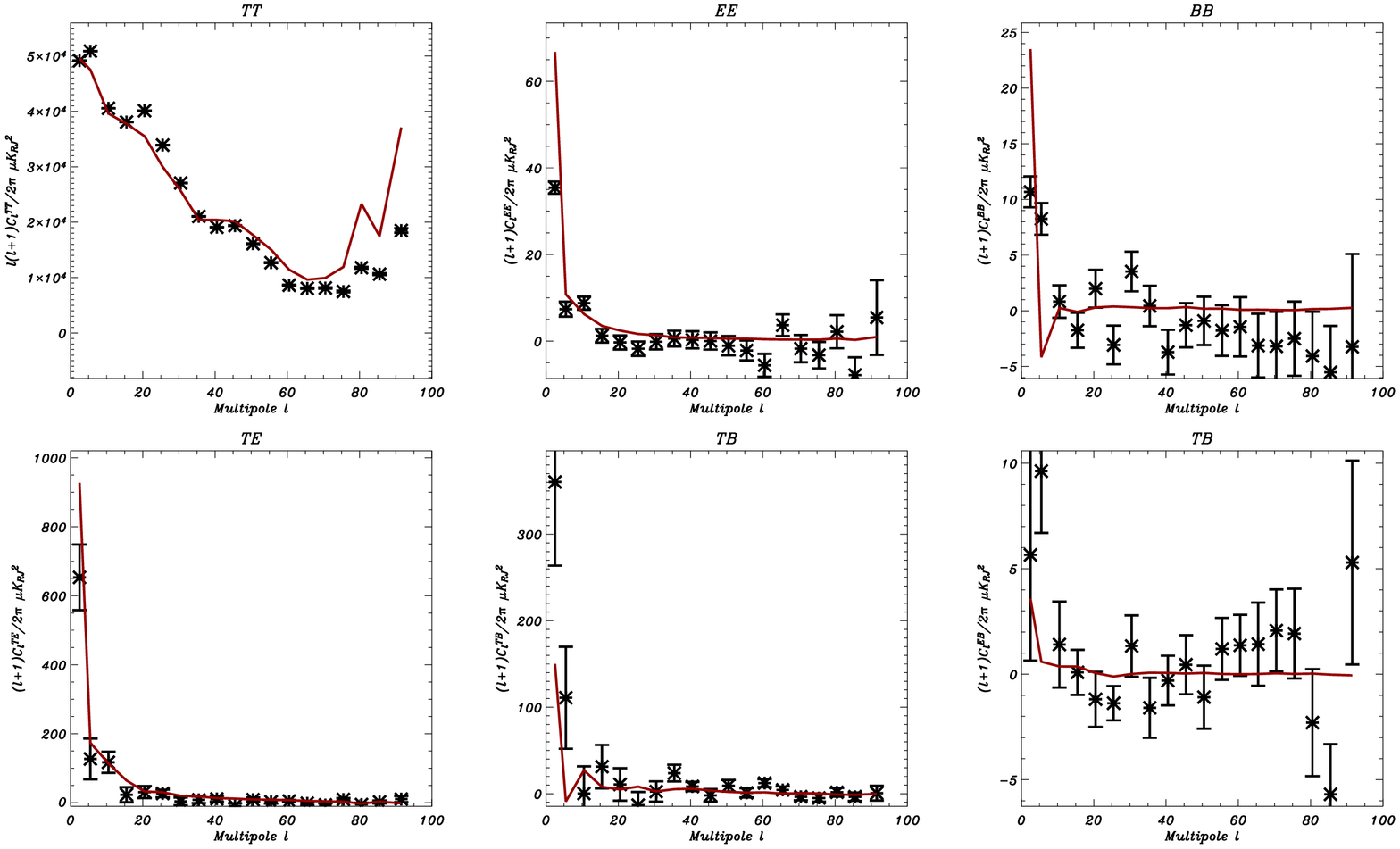}\caption{Clockwise from top left : power spectra $C^{TT}_l$, $C^{EE}_l$, $C^{BB}_l$, $C^{TE}_l$, $C^{TB}_l$, $C^{EB}_l$ at 353 GHz computed from Archeops data \emph{(black)} and the model of thermal dust emission with BSS magnetic field for the best fit model parameters \emph{(red)} without applying Galactic cut.\label{spec_dust_gp}}
\end{figure}

\indent Using the best fit parameters obtained for the Galactic
emissions models we computed maps and power spectra in temperature and polarization for synchrotron and dust thermal
emission. We compare them to maps and power spectra built respectively using polarized
WMAP and ARCHEOPS data. Like represented in the Figure~\ref{spec_wmap_gp} the synchrotron emission model is efficient to
reproduce the global feature of the data in polarization. We show on Figure~\ref{spec_dust_gp} the angular power spectra computed from the Archeops data at 353 GHz and the
thermal dust model using the method presented in~\cite{ponthieu2005}. Our model efficient to
reproduce the features of the spectra at all scales. 

\begin{figure}
\centering
\includegraphics[height=9cm,width=13cm]{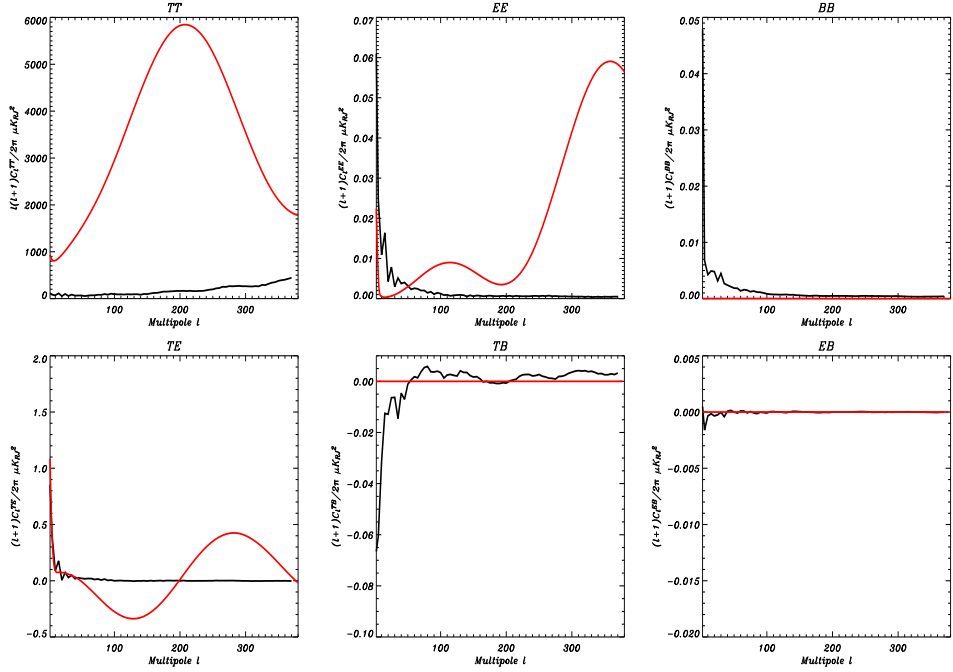}
\caption{Clockwise from top left : power spectra $C^{TT}_l$, $C^{EE}_l$, $C^{BB}_l$, $C^{TE}_l$, $C^{TB}_l$, $C^{EB}_l$ at 143 GHz for $|b|<15^{\circ}$ (see text for details).\label{fig:spect_cmb_for}}
\end{figure}

\indent From the above best fit parameters we estimate the contamination of the CMB PLANCK data by the
polarized galactic emissions. Figure~\ref{fig:spect_cmb_for}
shows the temperature and polarization power spectra at 143 GHz for
the CMB\footnote{We simulate CMB assuming cosmological parameters for a model $\Lambda$CDM like proposed in~\cite{komatsu} with a ratio tensor-scalar of 0.03.} (red)
and the Galactic foreground emissions applying a Galactic cut of 
$|b|<15^{\circ}$. The residual foreground contamination seems to be
weak but for the BB-modes for which an accurate foreground substraction is extremely important for
the detection of the primordial gravitational waves.

\section{Conclusions}

\indent We propose in this study consistent models of the main
Galactic polarized emissions based on a 3D modelisation of the
Galaxy. By comparison with preexisting data we are able to give
consistent constraints
on the parameters of the synchrotron and dust thermal emissions models
compatibles with thus appear in the literature. From this we build map
and power spectra enable to reproduce the features of the data at
various frequencies. Using a rough mask we then estimate the residual
contamination due to these foregrounds on the assumed CMB PLANCK data.


\end{document}